\documentclass[a4paper]{jpconf}

\usepackage{graphicx}
\usepackage{epsf,epsfig,amssymb,amsmath}
\usepackage{epstopdf}

\usepackage{color}

\newcommand{\be}{\begin{equation}}
\newcommand{\ee}{\end{equation}}
\newcommand{\bea}{\begin{eqnarray}}
\newcommand{\eea}{\end{eqnarray}}

\newcommand{\gton}{\mathrel{\lower.9ex \hbox{$\stackrel{\displaystyle 
>}{\sim}$}}} 
\newcommand{\lton}{\mathrel{\lower.9ex \hbox{$\stackrel{\displaystyle 
<}{\sim}$}}}

\newcommand{\vp}{{\bf p}}

\newcommand{\myell}{{\ell}}
\newcommand{\feq}{f^{\rm eq}}

\begin{document}
\title{Self-consistent Cooper-Frye freeze-out of a viscous fluid to particles}

\author{Zack Wolff and Denes Molnar}

\address{Department of Physics and Astronomy, Purdue University, 525 Northwestern Ave, West Lafayette, IN 47901, USA}

\ead{zwolff@purdue.edu}

\begin{abstract}
Comparing hydrodynamic simulations to heavy-ion data inevitably requires the conversion of the fluid to particles. This conversion, typically done in the Cooper-Frye formalism, is ambiguous for viscous fluids. We compute self-consistent phase space corrections by solving the linearized Boltzmann equation and contrast the solutions to those obtained using the ad-hoc ``democratic Grad'' ansatz typically employed in the literature where coefficients are independent of particle dynamics. Solutions are calculated analytically for a massless gas and numerically for both a pion-nucleon gas and for the general case of a hadron resonance gas. We find that the momentum dependence of the corrections in all systems investigated is best fit by a power close to 3/2 rather than the typically used quadratic ansatz. The effects on harmonic flow coefficients $v_2$ and $v_4$ are substantial, and should be taken into account when extracting medium properties from experimental data. $\quad$ PACS: 12.38.Mh, 24.10.Lx 24.10.Nz, 25.75.Ld, 51.10.+y

\end{abstract}

\section{Introduction}
The most widely used framework for describing the early stages of a heavy-ion collision is relativistic hydrodynamics\cite{hydro}. Using hydrodynamics requires one to convert the evolving fluid into a particle description in order to calculate experimental observables such as particle spectra and flow coefficients. This ``particlization''\cite{particlization} is usually done using the Cooper-Frye\cite{Cooper-Frye} formalism to calculate the particles emitted from a drop of the fluid on a constant temperature hypersurface. This requires knowledge of the distribution function of each particle species. If the fluid is in local thermal equilibrium, i.e. an ideal fluid, then one can uniquely calculate these distributions from the hydrodynamic variables. If the fluid contains dissipative effects such as those arising in a viscous fluid, infinitely many of these particle distributions will match the hydrodynamic fields. Other theory input is required to uniquely and self-consistently determine the particle distributions.

This ambiguity in the viscous particle distributions is typically ignored and in practice, corrections to the ideal thermal distributions are assumed to be quadratic in momentum. In addition to this quadratic ``Grad ansatz'', the distributions are also usually taken to be independent of the particle collision rates which keep the hadron gas near equilibrium. In this work we resolve this ambiguity by calculating the distributions self-consistently by taking reaction rates of the hadrons into account using the Boltzmann equation at particlization. We first calculate the distributions under the ``Grad ansatz'' which assumes the corrections are quadratic in momentum and then later relax this constraint. Many of the calculational details absent here can be found in \cite{preprint}.

\section{Phase space corrections from linearized transport}

The Cooper-Frye particlization prescription requires as input distribution functions for each particle species i which are calculable from the hydrodynamic fields by inverting $T^{\mu\nu}(x) \equiv \sum\limits_i \int\limits \frac{d^3p}{E} p^\mu p^\nu f_i(x,\vp)$. For ideal fluids in local equilibrium with known chemical potentials, the conversion is well defined since knowledge of the energy density in the energy-momentum tensor is enough to calculate the temperature needed for the thermal particle distributions assuming the equation of state is a known input of the hydrodynamic simulation.  

However, if the fluid has nonzero viscosity, it is not in perfect local equilibrium and the energy-momentum tensor acquires non-ideal corrections. In addition to the previously known ideal component, one must now solve for the viscous corrections to the distributions by inverting

\be
\delta T^{\mu\nu} = \pi^{\mu\nu} + \Pi (u^\mu u^\nu - g^{\mu\nu}) \equiv \sum\limits_i \int\limits \frac{d^3 p}{E} p^\mu p^\nu \delta f_i(x,\vp)
\ee

Without additional theory constraint on the functional form of $\delta f_i$, infinitely many functions are possible and one is left to ``guess'' at the correct distributions. If one takes corrections from shear only and ignores bulk viscosity, one common choice that satisfies the above constraint is the ``democratic Grad''\cite{democratic} ansatz which assumes corrections quadratic in momentum 

\be
\phi_i^{\rm dem}(x,\vp) = \frac{\pi^{\mu\nu}(x)p_\mu p_\nu}{2[e(x)+p(x)]T^2(x)} \ , \qquad \left(\delta f_i \equiv \phi_i f_i^{eq}\right)
\ee

This simple choice has no dependence on the microscopic dynamics of the particles and ignores the fact that species which interact more should be closer to equilibrium then those that rarely scatter off other particles. In contrast to this democratic ansatz, we calculate the viscous corrections self-consistently from linearized covariant transport theory where evolution is governed by the Boltzmann equation

\be
p^\mu \partial_\mu f_{i}(x,\vp) = S_i(x,\vp) + \sum\limits_{jk\myell} 
C^{ij\to k\myell}[f_i,f_j,f_k,f_\myell](x, \vp)
\ee
with S the source term specifying the initial conditions and collision terms here taken as

\be
C^{ij\to k\myell}[f_i,f_j,f_k,f_\myell](x,\vp_1) 
\equiv \int\limits_2 \!\!\!\!\int\limits_3 \!\!\!\!\int\limits_4
\left(\frac{g_i g_j}{g_k g_\myell} f_{3k} f_{4\myell} - f_{1i} f_{2j}\right)
\, \bar W_{12\to 34}^{ij\to k\myell}  \, \delta^4(12 - 34)
\ee
with shorthands
$\int\limits_a \equiv \int d^3p_a / (2 E_a)$, 
$f_{ai} \equiv f_i(x,\vp_a)$, and
$\delta^4(ab - cd) \equiv \delta^4(p_a + p_b - p_c - p_d)$. and $g_i$ the internal degrees of freedom.

If the viscous corrections are small, one can linearize the Boltzmann equation in these corrections $\delta f_i$. Most systems quickly relax to a solution of these equations which is governed by gradients of the equilibrium distribution, i.e. the Navier-Stokes regime where derivatives of $\delta f_i$ are higher order corrections. The tensor structure of this equation imposes the form of the solutions\cite{preprint}
\be
\phi_i(x,\vp)
   = \chi_i(|\tilde\vp|) P^{\mu\nu} X_{\mu\nu} 
\ee
with $\tilde\vp$ being momentum over temperature in the fluid rest frame and tensor shorthands 
\be
P_a^{\mu\nu} \equiv \frac{1}{T^2}\left[
        \Delta^\mu_\alpha \Delta^\nu_\beta p_a^\alpha p_a^\beta - 
          \frac{1}{3}\Delta^{\mu\nu} (\Delta_{\alpha\beta} p_a^\alpha p_a^\beta)
        \right]
\ , \ \
X^{\mu\nu} \equiv \frac{\sigma^{\mu\nu}}{T} 
 = \frac{\pi^{\mu\nu}_{NS}}{\eta_s T}\ , \ \ \Delta^{\mu\nu} \equiv g^{\mu\nu}-u^\mu u^\nu
\ee
with $\sigma{\mu\nu} \equiv \nabla^\mu u^\nu + \nabla^\nu u^\mu - \frac{2}{3}\Delta^{\mu\nu} (\partial u)$ The only freedom left in the distribution function corrections is a function of dimensionless momentum $\chi_i(|\tilde\vp|)$ for each particle species. One can show that solving the original transport equation is equivalent to extremizing the functional\cite{AMY}

\bea
Q[\chi] =  \frac{1}{2T^2} \sum\limits_i \int\limits_1 P_1 \cdot P_1 \feq_{1i} \chi_{1i} &+& \frac{1}{2T^4} \! \sum\limits_{ijk\myell}
                        \int\limits_1\!\!\!\!\int\limits_2\!\!\!\!
                        \int\limits_3\!\!\!\!\int\limits_4
           \!\!\feq_{1i} \feq_{2j} \, \bar W_{12\to 34}^{ij\to k\myell}\,
           \delta^4(12-34) \cdot
\nonumber\\
         && (  \chi_{3k} P_3 \cdot P_1 + \chi_{4\myell} P_4 \cdot P_1
              - \chi_{1i} P_1 \cdot P_1 - \chi_{2j} P_2 \cdot P_1) \chi_{1i}
\eea
and the extremal value of Q gives the shear viscosity of the mixture
\be
\eta_s = \frac{2}{15T^3} \sum\limits_i \int \frac{d^3 p }{E} p^4 \feq_i \chi_i \Rightarrow \eta_s = \frac{8}{5} Q_{max} T^3
\ee
In contrast, the Grad ansatz corresponds to giving all particles the same coefficient
$\chi_i^{\rm dem} = \frac{\eta_s T}{2(e+p)} = \frac{\eta_s}{2s}$  independent of momentum and particle species.

\section{Results for massless two-component system}

In this section we consider the so-called Grad ansatz where $\chi_i(|\tilde\vp|) = \chi_i^{Grad} = const$   such that the viscous corrections $\phi_i$ are quadratic in momentum. For a two-component system of massless particles evolving with elastic two-body interactions with simple isotropic, energy-independent cross sections $\sigma_{AA}$, $\sigma_{BB}$, and $\sigma_{AB}$ in the Grad approximation we can solve the functional method analytically\cite{preprint} and get

\bea
\chi_A^{Grad} &=& \frac{3LT}{20}
         \frac{5K_{B(B)} + 7K_{B(A)} + 2K_{A(B)}}
              {K_{A(A)} [5 K_{B(B)} + 7 K_{B(A)}]
               + K_{A(B)} [9K_{B(A)} + 7 K_{B(B)}]}
\\
\chi_B^{Grad} &=&\frac{3LT}{20}
         \frac{5K_{A(A)} + 7K_{A(B)} + 2K_{B(A)}}
              {K_{B(B)} [5 K_{A(A)} + 7 K_{A(B)}]
               + K_{B(A)} [9 K_{A(B)} + 7 K_{A(A)}]} \ .
\eea
where $K_{i(j)} = L n_j \sigma_{ij}$ are partial inverse Knudsen numbers for species $i$ scattering off species $j$ and $L$ is the characteristic length scale of gradients in the system. 

These results were derived using the linearized transport equation assuming small deviations from local equilibrium which corresponds to the Navier-Stokes limit of hydrodynamics where the system has relaxed to a solution governed by gradients of the hydrodynamic variables. In heavy-ion collisions, this relaxation to equilibrium has to compete with expansion diluting the system. We now test how well this Navier-Stokes solution works for expanding systems with a two-component massless system undergoing boost-invariant 0+1D Bjorken expansion as in \cite{Navier-Stokes}. The system starts out in equilibrium at $\tau_0$ but expansion causes viscous corrections to quickly develop. These corrections can be expressed simply in terms of the partial sheer stresses of the two species. If we assume dissipative corrections quadratic in momentum then our $\chi_i$'s are constant $ \equiv c_i$. In the late-time Navier Stokes regime, linearized kinetic theory predicts 

\be
\frac{c_B}{c_A} = \frac{5 K_{A} + 2(K_{A(B)} + K_{B(A)})}
                       { 5K_B + 2(K_{A(B)}+K_{B(A)})} \qquad\qquad
(K_i \equiv \sum\limits_j K_{i(j)}) \ .
\label{lingradcAcB}
\ee
While the ``democratic Grad'' approach gives $c_i = 1$ for all species, so $c_B / c_A = 1$.

\begin{figure}[h]
\begin{minipage}{9cm}
\includegraphics[width=9cm]{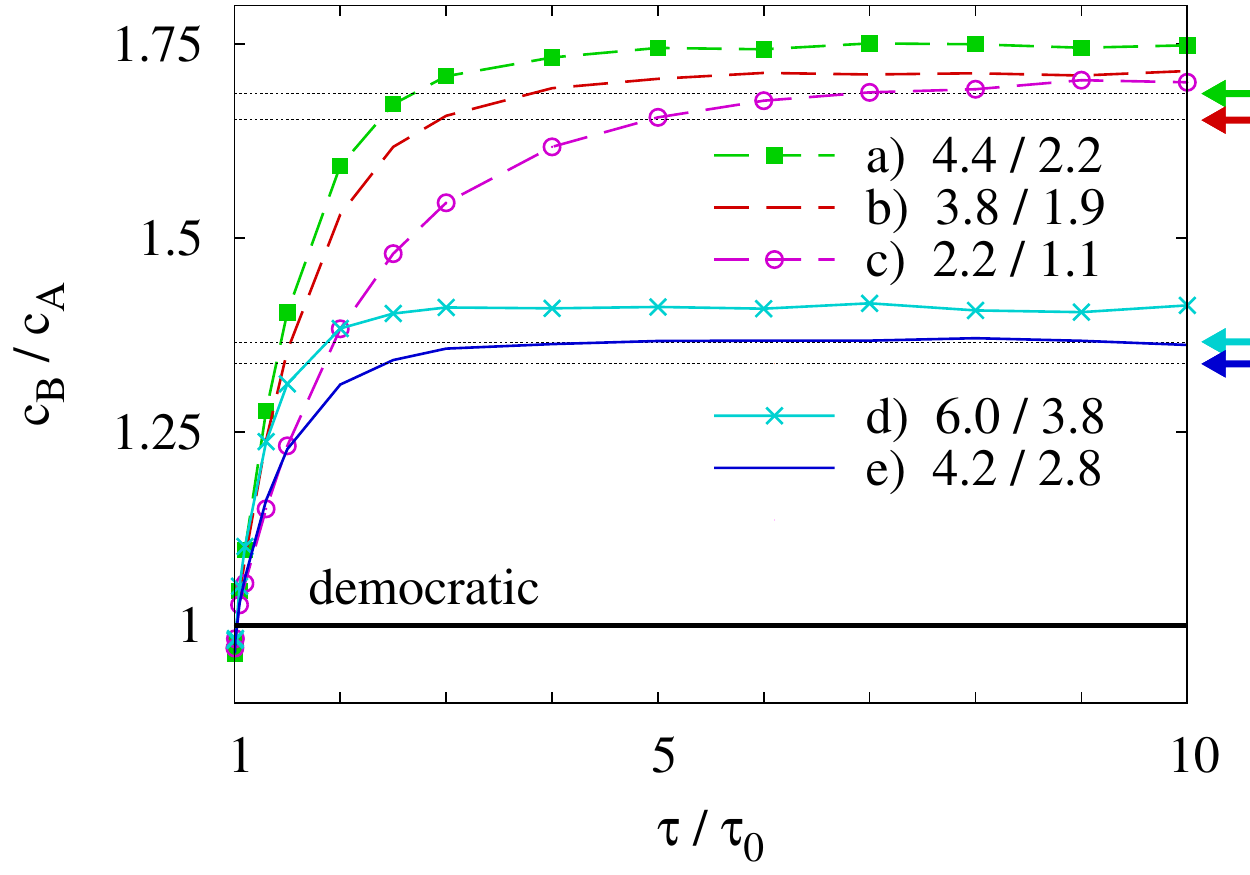}
\end{minipage}
\hspace{0cm}
\begin{minipage}{8cm}
\footnotesize
\centering
\vspace{-.5cm}
\begin{tabular}{|c|c|c|c|c|}
\hline
Scenario & $K_A$ & $K_B$ & 
$n_A : n_B$ & $\sigma_{AA} : \sigma_{AB} : \sigma_{BB}$ \cr
\hline
\hline
a) & 4.4 & 2.2 & 3 : 1 & 20 : 10   :  5 \cr
b) & 3.8 & 1.9 & 2 : 2 & 20 : 10   :  5 \cr
c) & 2.2 & 1.1 & 2 : 2 & 12 :  6   :  3 \cr
d) & 6   & 3.8 & 1 : 3 & 24 : 24   : 12 \cr
e) & 4.2 & 2.8 & 2 : 2 & 20 : 13.3 : 8.89 \cr
\end{tabular}
\caption{\label{cAcB}Ratio of viscous corrections as a function of dimensionless proper time for a massless two-component system undergoing 0+1D Bjorken expansion. Thin dotted lines and arrows refer to the expectation based on the linear transport equation and the numbers give the ratio of inverse Knudsen numbers, $K_A / K_B$ as shown in the table.}
\end{minipage}
\end{figure}

Fig. \ref{cAcB} compares these two extreme cases to fully nonlinear transport solutions from Molnar's Parton Cascade (MPC) \cite{mpc}. The five different scenarios listed in the table of Fig. \ref{cAcB}, all keeping species A closer to equilibrium than species B, were simulated with MPC. All scenarios start in equilibrium where $c_B / c_A = 1$, but then expansion forces them out of equilibrium. At late times the dissipative corrections predicted by the linear transport equation are correct within 10\% while the commonly used ``democratic Grad'' ansatz fails to account for the species dependence of corrections. This gives confidence that the Navier-Stokes regime described by the linearized transport equation can be appropriate to use for heavy-ion expanding systems.

\section{Results for massive two-component system: pion-nucleon gas}

In \cite{preprint} it was shown that a nonrelativistic particle in the dynamic Grad approach described in the previous section receives a correction inversely proportional to the square root of its mass which reproduces the canonical viscosity expression in this limit:
\be
\chi^{Grad} = \frac{5\sqrt{\pi}}{32}\sqrt{\frac{T}{m}} \frac{T}{n\sigma_{TOT}} 
\quad
\Rightarrow \quad
\eta_s^{Grad} = \frac{5\sqrt{\pi}}{16} \frac{\sqrt{m T}}{\sigma_{TOT}} \ ,
\label{etas_NR_Grad}
\ee
In this section, we now consider a more realistic pion-nucleon system with relativistic kinematics. Lumping isospin states and antiparticles into a single species, this is a two-component system with $m_\pi = 0.14$ GeV, $g_\pi = 3$, $m_N = 0.94$ GeV, $g_N = 4$. For temperatures $120$~MeV $\lton T \lton 165$~MeV of interest, we approximate the two-body cross sections with constant, energy-independent, effective values $\sigma^{eff}_{\pi\pi} = 30$~mb, $\sigma^{eff}_{\pi N} = 50$~mb, 
and $\sigma^{eff}_{NN} = 20$~mb which reproduce the pion and proton relaxation times in \cite{Prakash}. Note that pions are much more prevalent at these temperatures so the nucleon-nucleon scattering is mostly irrelevant. The ratio of pion to proton coefficients in this temperature range $c_\pi / c_N \sim 2$ turns out to be near the nonrelativistic calculation predictions, even though pions are relativistic. This means nucleons are about twice as close to equilibrium as pions of the same momentum.
This pion-proton difference will be manifest in identified particle observables if the self-consistent, species dependent viscous corrections are used in Cooper-Frye freeze-out. 

To find the effect on pion and proton elliptic flow in heavy-ion collisions we run a hydrodynamic simulation of a RHIC Au+Au collision at $\sqrt{s_{NN}}=200$~GeV with impact parameter $b=7$~fm. The calculation was done with AZHYDRO\cite{AZHYDRO} version 0.2p2 which is a 2+1D hydro code with longitudinal boost invariance and the fairly up-to-date s95-p1 equation of state parametrization\cite{EOSs95p1} that matches lattice QCD results to a hadron resonance gas. Since AZHYDRO does not contain dissipation, the shear stress is estimated on the conversion hypersurface from gradients of the flow using the Navier-Stokes estimate: $\pi^{\mu\nu} = \eta_s \sigma^{\mu\nu}$. 

\begin{figure}[h]
\begin{minipage}{9cm}
\includegraphics[width=9cm]{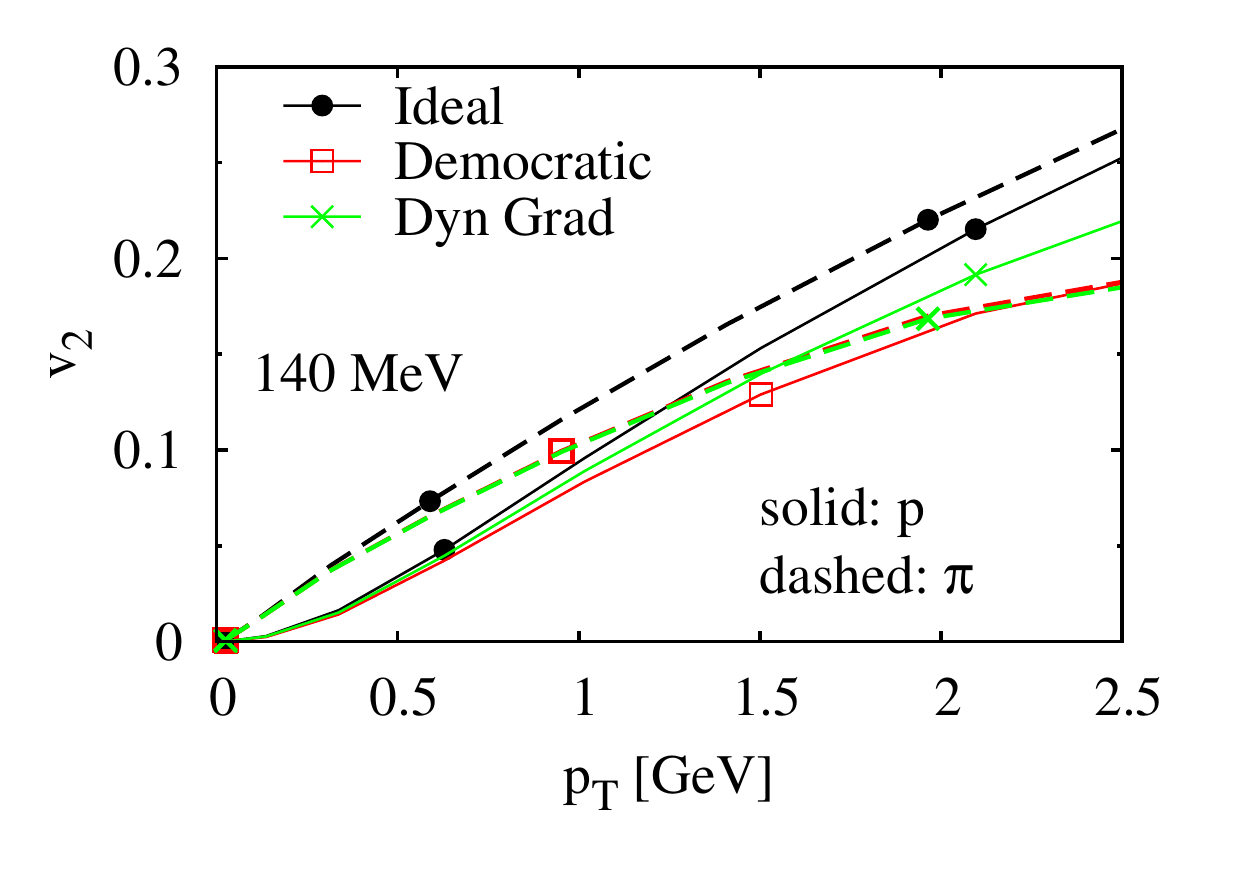}
\end{minipage}
\hspace{-1cm}
\begin{minipage}{9cm}
\includegraphics[width=9cm]{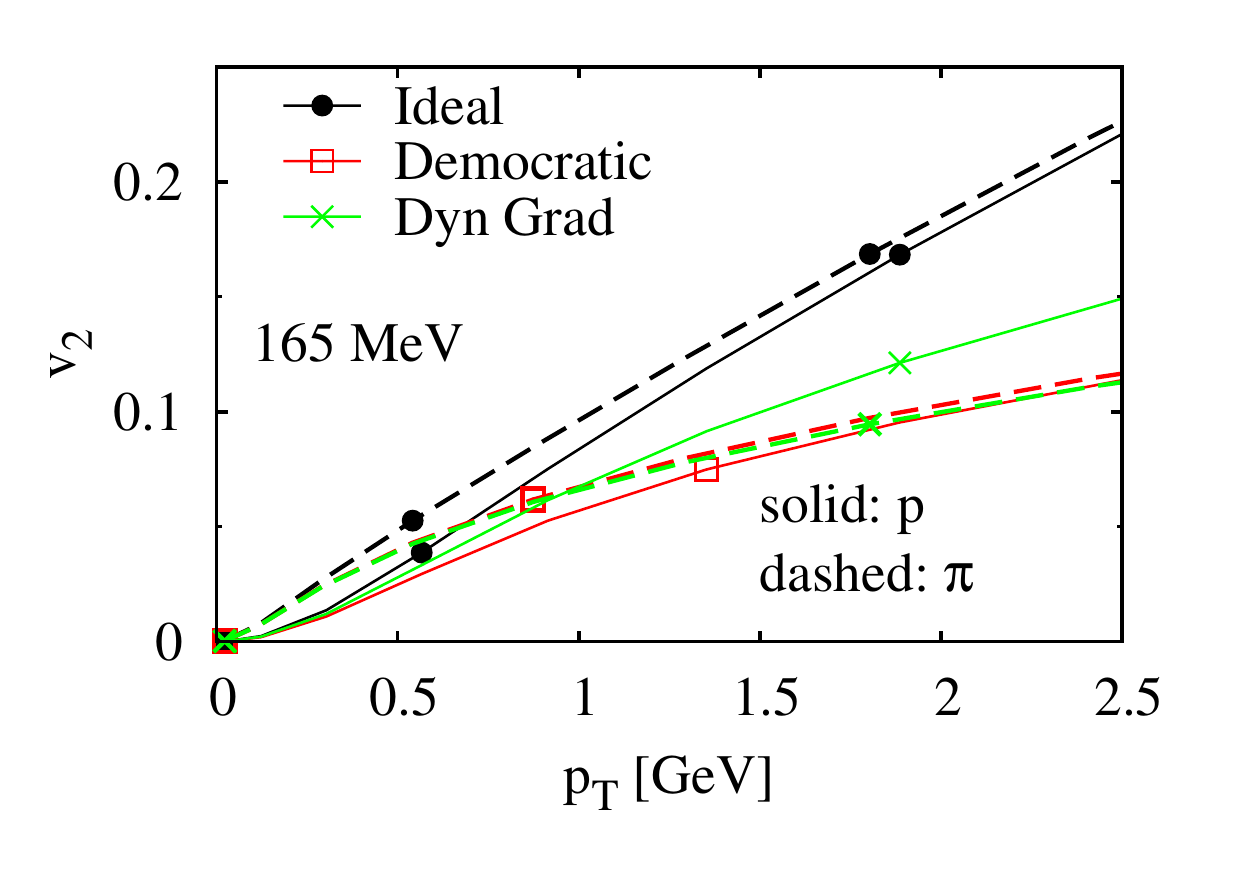}
\end{minipage}
\vspace{-0.6cm}
\caption{Differential elliptic flow $v_2(p_T)$ of pions and protons from the AZHYDRO simulation described in the text for $T_{conv}=140$ MeV (left) and $T_{conv}=165$ (right). Pions are shown in dashed lines and protons in solid. The commonly used ``democratic Grad'' approach (open boxes) is compared to self-consistent shear corrections (crosses) computed for a pion-nucleon gas from linearized kinetic theory. Results with no viscous corrections ($\delta f = 0$) (filled circles) are also shown for reference.}
\label{v2-pi_p}
\end{figure}
Fig. \ref{v2-pi_p} shows differential elliptic flow for pions and protons for freeze-out at $T_{conv} = 140$ MeV (left) and $T_{conv} = 165$ MeV (right). In the ideal case (filled circles) one can already see the mass ordering of the pion and proton $v_2$ curves characteristic of hydrodynamic flow. When dissipative effects are calculated with the commonly used democratic Grad ansatz (open boxes), the mass ordering persists but elliptic flow is strongly suppressed by viscous effects at moderate momentum. In contrast, self-consistent species-dependent freeze-out (crosses) causes a clear pion-proton elliptic flow splitting at the higher values of $p_T$ with proton $v_2$ being $\sim 30\%$ higher than that of the pion. The qualitative picture is similar at the two temperatures, with the noticeable difference of reduced viscous effects at 140 MeV due to smaller flow gradients at later ``times''.

\section{Results for a multicomponent hadron gas}

In the previous section, self-consistent viscous corrections were calculated for a pion-nucleon gas. While these are the main stable particles whose flow is measured in experiments (along with kaons), this is a rather large simplification of the system expected to be present after hadronization in a heavy-ion collision.  The hadron gas is made up of many more species and resonances whose scattering rates with all other species should be taken into account in the calculation. Here we use two different simplifying scenarios to categorize the cross sections present in a full multicomponent hadron gas. In the first scenario we give all species the same constant, energy-independent scattering cross section with every other species, namely 30mb similar to \cite{Denicol}. In the second scenario we use the additive quark model (AQM)\cite{AQM} to motivate cross sections which scale as the product of the number of quarks in the two hadrons, namely the meson-baryon cross sections will be in ratios of $\sigma_{MM} : \sigma_{MB} : \sigma_{BB} = 4:6:9 = 30:45:67.5$ mb. In both cases we consider only elastic $ij\to ij$ channels (allowing for $i=j$). As before, we group members of the same isospin multiplet and antiparticles into a single species with degeneracy factors adjusted accordingly. The calculation was done including all hadrons up to $m=1.672$ GeV, i.e. the $\Omega(1672)$, which translates into 49 effective species.

\begin{figure}[h]
\begin{minipage}{9cm}
\includegraphics[width=9cm]{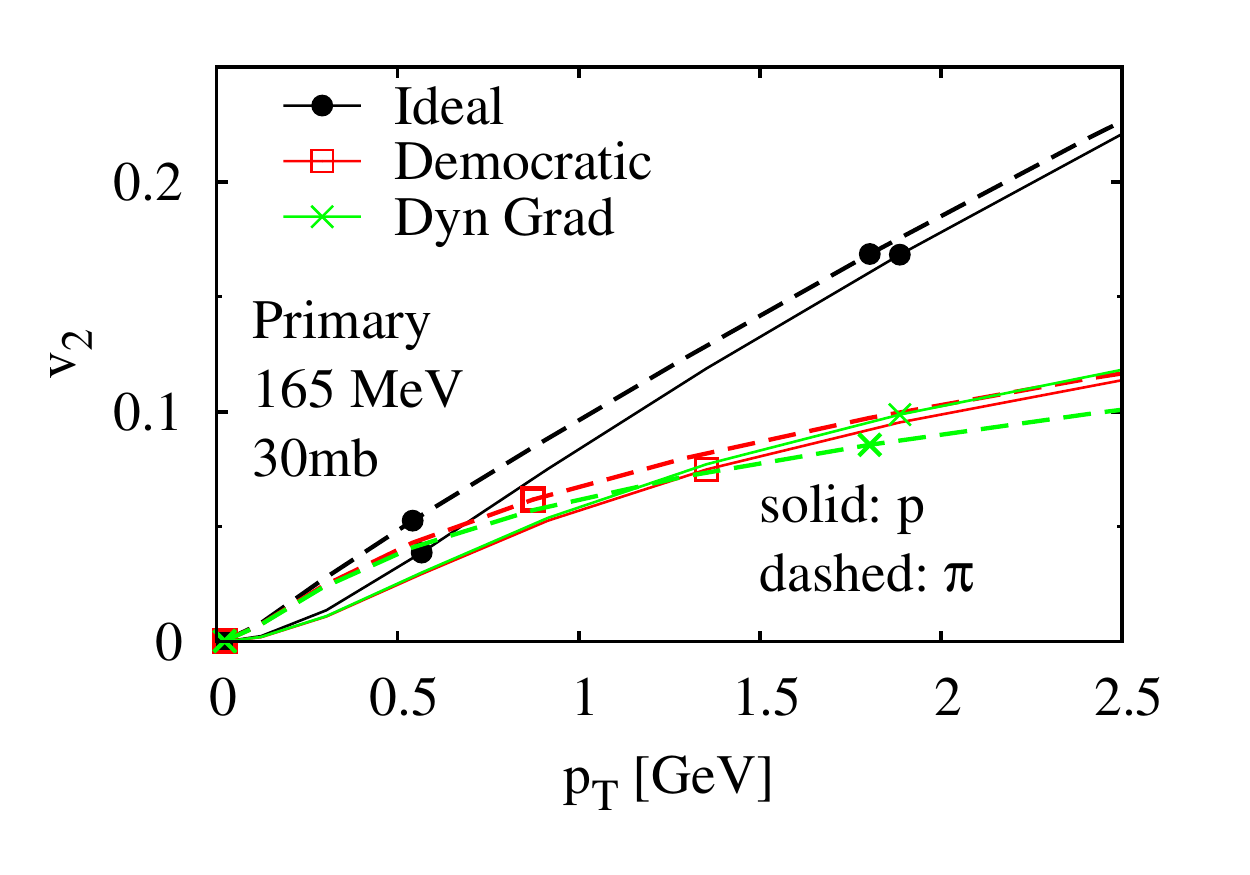}
\end{minipage}
\hspace{-1cm}
\begin{minipage}{9cm}
\includegraphics[width=9cm]{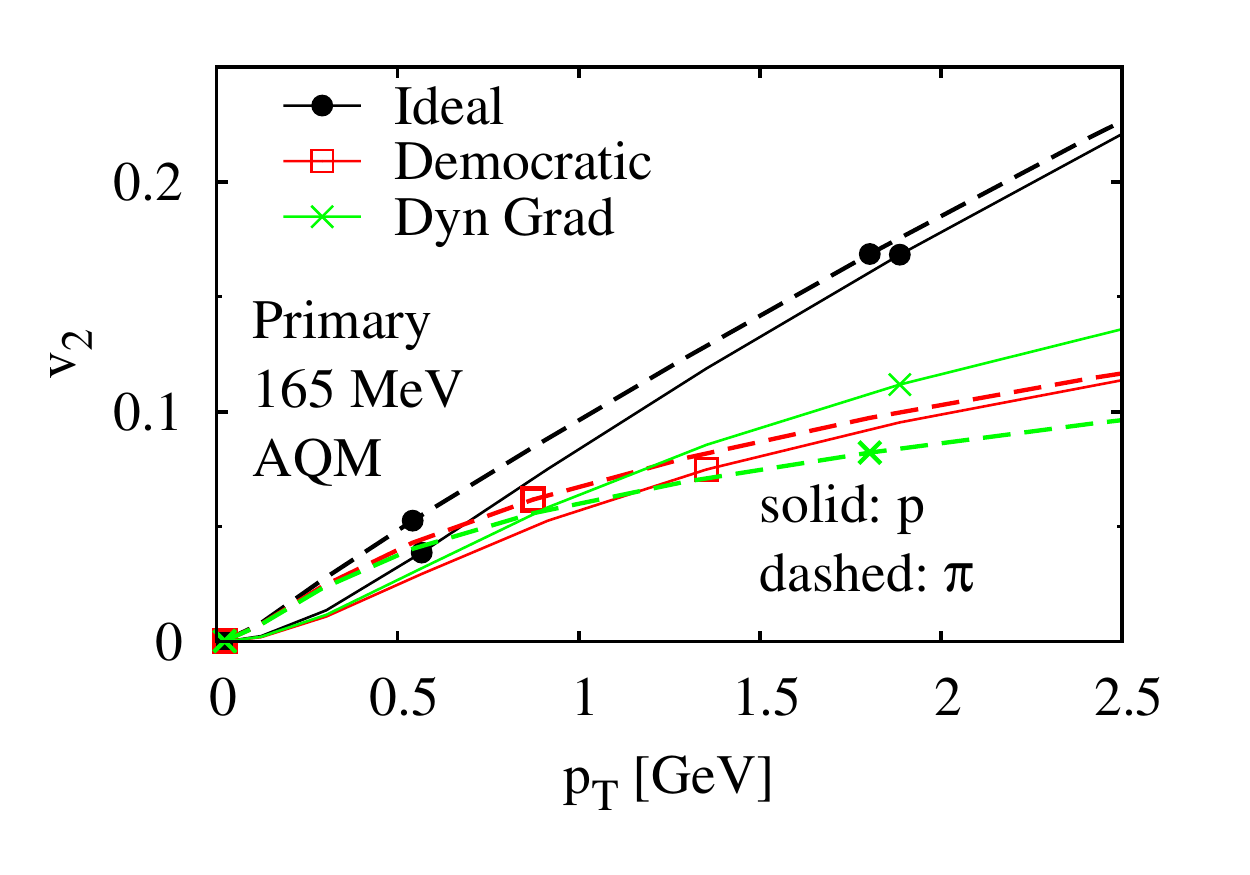}
\end{minipage}
\vspace{-0.6cm}
\caption{\label{primary}Same as Fig. \ref{v2-pi_p} but the self-consistent species-dependent coefficients were calculated for the multicomponent hadron gas with 49 effective species with decoupling temperature $T_{conv} = 165$ MeV. The left plot shows the first scenario described in the text where all species have a constant 30 mb cross section with each other species. The right plot shows the second scenario motivated by the additive quark model (AQM) with $\sigma_{MM} : \sigma_{MB} : \sigma_{BB} = 30:45:67.5$ mb.}
\end{figure}

Fig. \ref{primary} shows pion and proton elliptic flow $v_2(p_T)$ in Au+Au at RHIC at b=7 fm from a calculation analogous to the pion-nucelon system in the previous section with particlization at $T_{conv} = 165$ MeV, except now the self-consistent calculation has been done for a multicomponent hadron gas. The left plot shows that with the same constant cross section between all species, the result is similar to the democratic case, as found previously in \cite{Denicol}.

In contrast, the right plot shows that if the more realistic hadron gas with AQM motivated cross sections is used, the results on pion and proton identified flows are noticeable. Again they cross at $p_T \sim 1$ GeV and behavior is roughly similar to the pion-nucleon gas. Though not shown, results are similar at a decoupling temperature $T_{conv} = 140$ MeV. 

\begin{figure}[h]
\begin{minipage}{9cm}
\includegraphics[width=9cm]{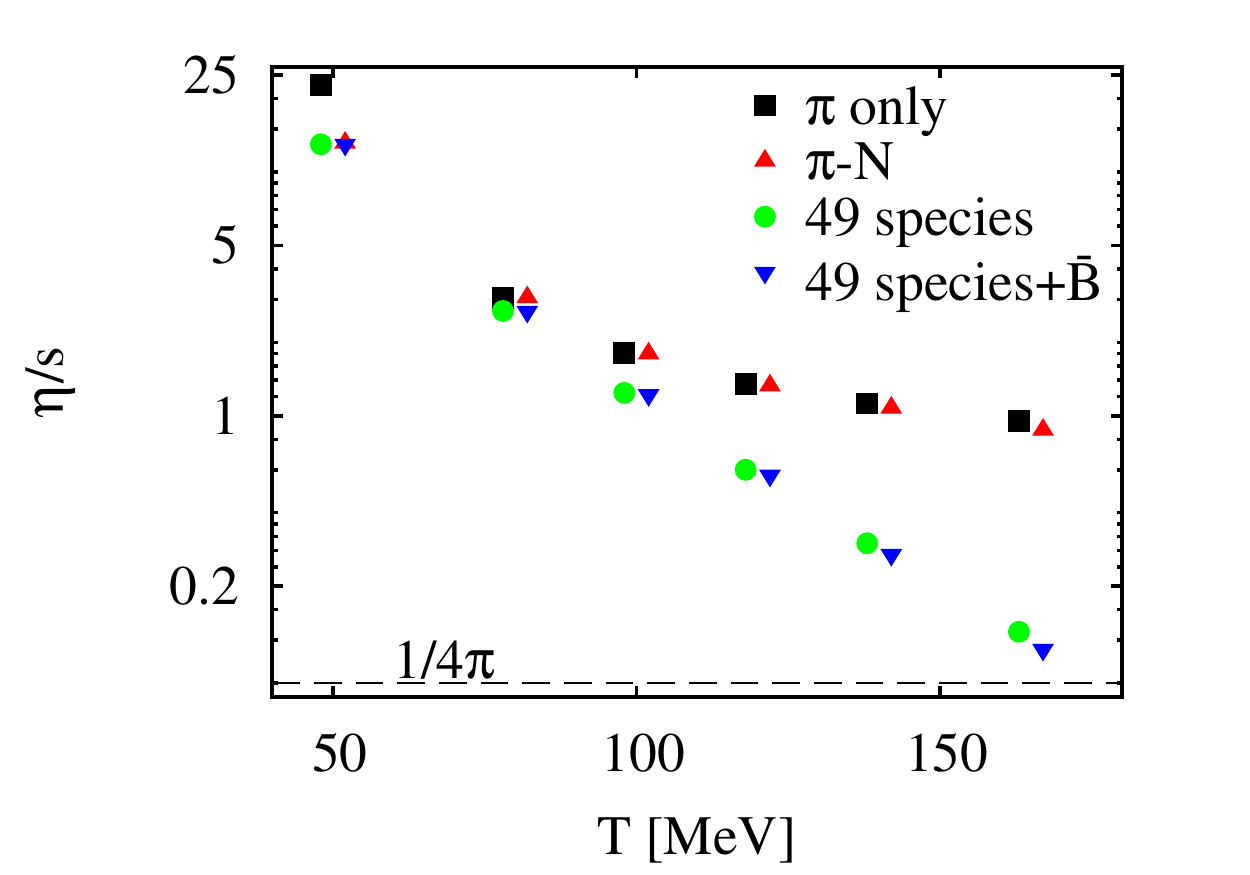}
\end{minipage}
\hspace{-1cm}
\begin{minipage}{7.5cm}
\vspace{-2cm}
\caption{\label{etaos}Shear viscosity to entropy density ratio calculated from the functional method described in the text for pions (squares), a pion-nucleon gas (triangles), the hadron gas without antiparticles (circles), and the full hadron gas (upside-down triangles).}
\end{minipage}
\end{figure}

As mentioned previously, the functional extremization approach used here automatically gives the shear viscosity of the gas after particlization and results for 4 systems are given in Fig. \ref{etaos}. As one can see, the value of the shear viscosity to entropy density $\frac{\eta}{s}$ of the hadron gas falls close to the proposed limit of $1/4\pi$ at 165 MeV due to the rapid rising of the entropy at high temperature.

\begin{figure}[h]
\begin{minipage}{9cm}
\includegraphics[width=9cm]{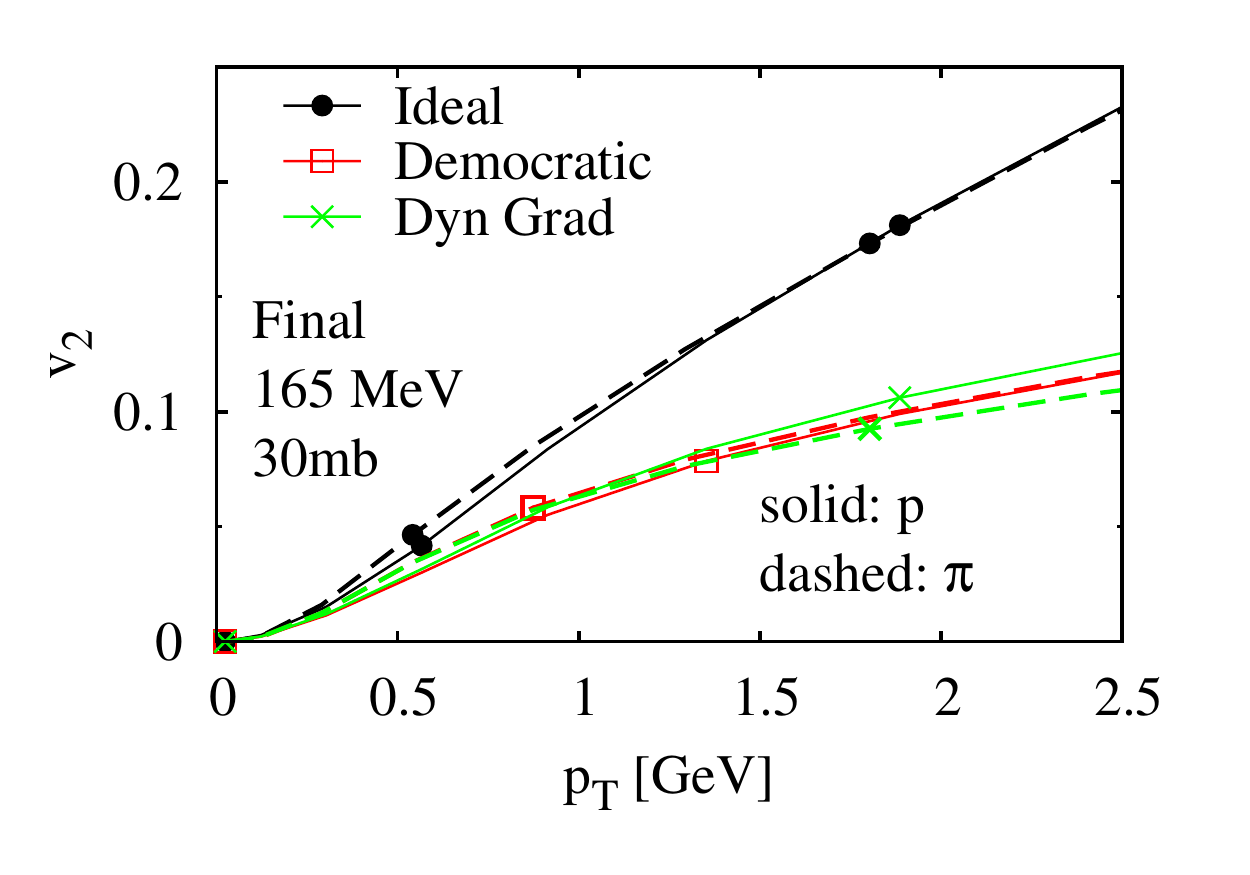}
\end{minipage}
\hspace{-1cm}
\begin{minipage}{9cm}
\includegraphics[width=9cm]{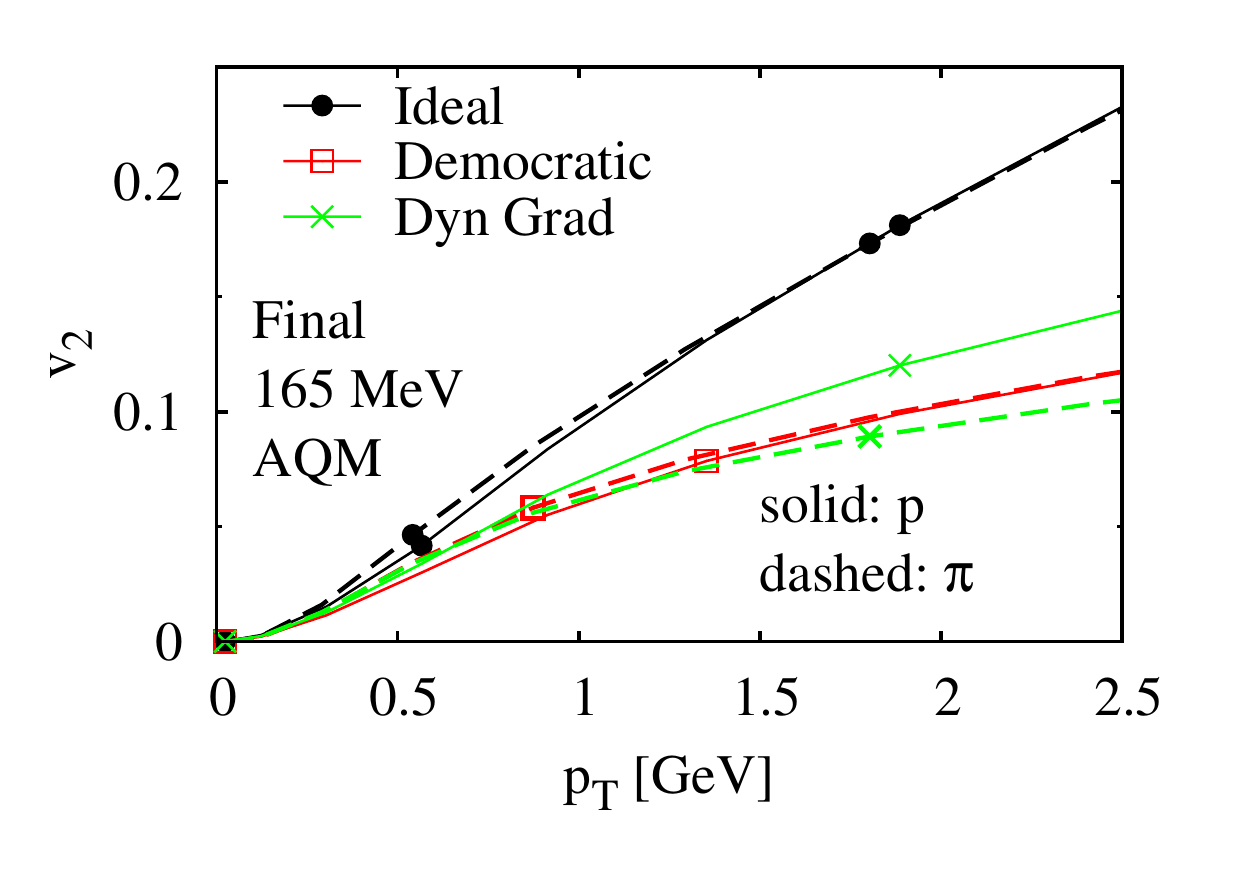}
\end{minipage}
\vspace{-0.6cm}
\caption{\label{final}Same as Fig. \ref{primary} except after feed down from resonance decays using the RESO code in the AZHYDRO package\cite{AZHYDRO}.}
\end{figure}

The Cooper-Frye prescription gives the momentum distribution of particles coming directly from the fluid. Whether one couples the hydrodynamic simulation to a hadron transport code, the so-called ``afterburner'' approach, or the particles are allowed to free stream to the detectors, one should allow all of the unstable resonances to decay to stable pions and protons and include these secondary particles in the spectra and elliptic flow calculations of these stable particles. This resonance feed down has not been calculated in Fig. \ref{primary}, but is included in Fig. \ref{final} below with the help of the AZHYDRO companion code RESO\cite{AZHYDRO2}. For ideal freeze-out ($\delta f$ =0), the democratic freeze-out, and the constant cross section scenario, the main effect of the resonance decays is reduction of the pion-proton splitting. At $T_{conv} = 165$ MeV, there is almost no difference in pion and proton elliptic flow (although for lower conversion temperatures, some difference does remain). Again in contrast to the other freeze-out scenarios, the self-consistent freeze-out based on the linear transport equation gives a measurable increase in proton elliptic flow, $\sim 30\%$ compared to pion flow around the higher momentum of around 2 GeV. This amount of splitting should be detectable in experiment and can be used to distinguish between different freeze-out distributions.

\section{Results for hadron gas with $\delta f_i \propto p^{3/2}$}

\begin{figure}[h]
\vspace{-0.5cm}
\begin{minipage}{9cm}
\includegraphics[width=9cm]{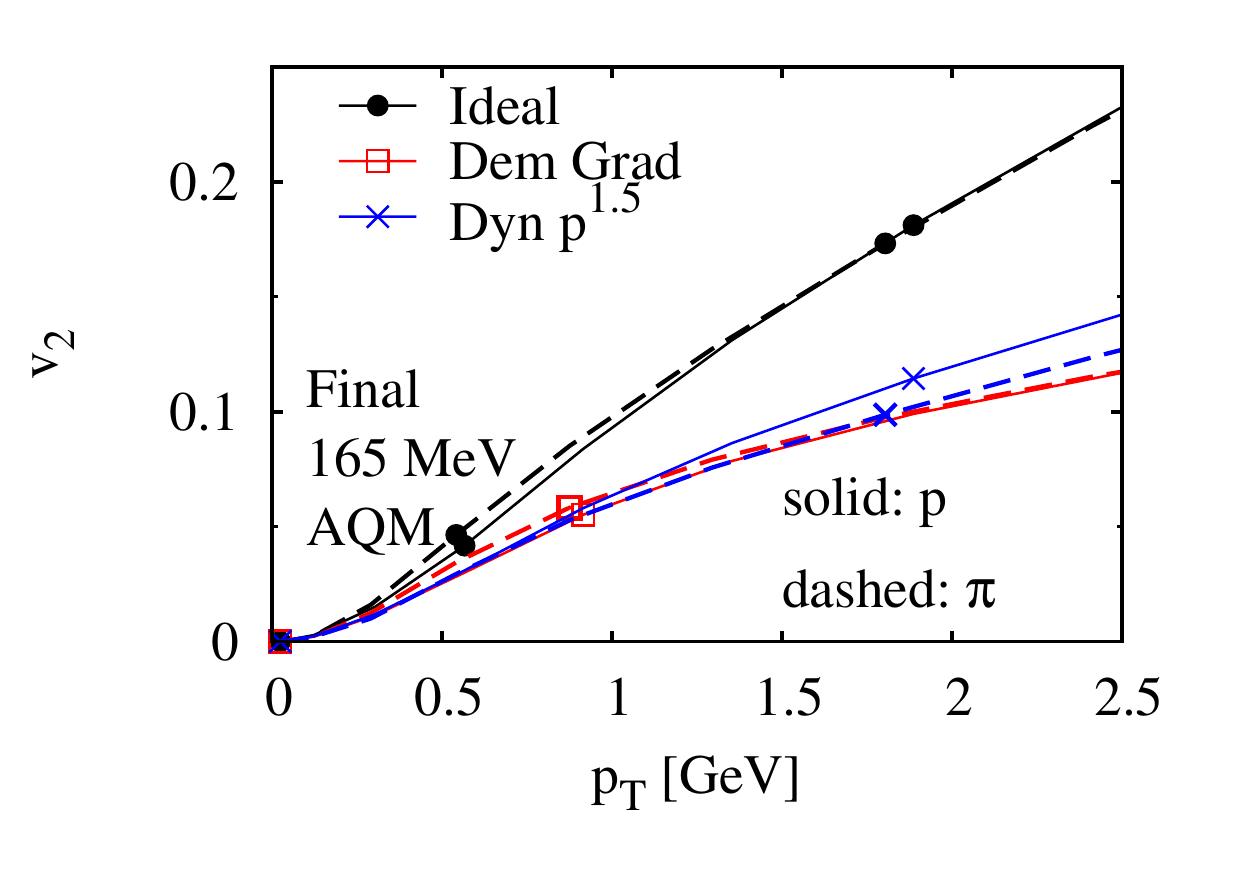}
\end{minipage}
\hspace{-0.5cm}
\begin{minipage}{9cm}
\includegraphics[width=9cm]{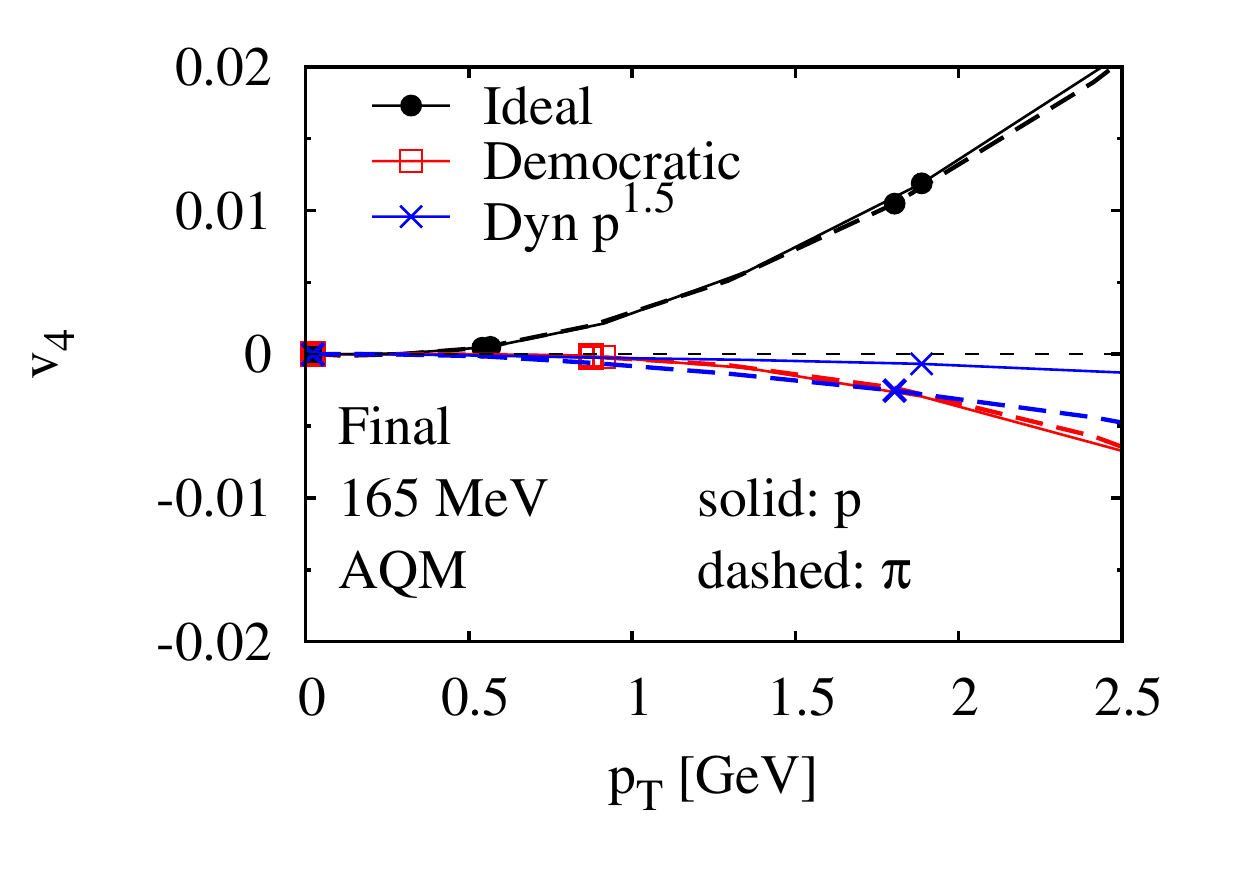}
\end{minipage}
\vspace{-0.6cm}
\caption{\label{pto3o2}Same as Fig. \ref{final} (right) except the self-consistent calculation here was done for corrections proportional to momentum to the 1.5 instead of quadratic for $v_2$ (left) and $v_4$ (right).}
\end{figure}

Up to this point, we have stayed with the commonly used assumption that dissipative corrections to phase space distributions from shear viscosity are quadratic in momentum. The quadratic dependence used so far was essentially a simple guess at the real solution, while our integral equations for $\chi_i$ still contain an unknown function of the magnitude of the momentum. A better method then ``guessing'' at the correct form of the solution would be to either calculate which simple functional form better extremizes the functional and/or compare the results for observables obtained from different choices of the functional form of the corrections. We chose to do the former and using our analytic solution to the massless problem calculated the single power that best maximizes the functional. The answer we came up with was near $\frac{3}{2}$, a value which has been found in other theoretical and phenomenological studies\cite{Dusling}\cite{Luzum}. We then chose this momentum to the $\frac{3}{2}$ power as our functional form of $\chi_i$ and again calculated what the results for pion and proton elliptic flow would be assuming hydrodynamic freeze-out to a hadron gas. The result after resonance decays is shown in the left plot of Fig. \ref{pto3o2}.

When the viscous corrections are calculated using a momentum to the $\frac{3}{2}$ power instead of quadratic, the elliptic flow of pions and protons comes closer together at higher momentum. Both pion and proton $v_2$ go above the democratic curves at high momentum do to the smaller viscous corrections there. This is compensated by the self-consistent curves being below the democratic ones at low momentum. The effects of this new momentum dependence on higher harmonics were also studied and the results for $v_4$ are shown in the right plot of Fig. \ref{pto3o2}.

The effect on identified particle $v_4$ is quite pronounced. The self-consistent calculation brings the proton up to near zero and actually positive in the quadratic case not pictured here. The effect on the sixth coefficient is less pronounced as both the quadratic and $p^{1.5}$ corrections have small deviations from the democratic case in this simulation and are thus not shown here.

\section{Summary}

Extracting accurate values for medium properties from heavy-ion data requires proper treatment of the viscous corrections to hadron distribution functions in the Cooper-Frye freeze-out whether one uses hydrodynamics or a hybrid hydrodynamics+transport simulation. Here we have calculated these corrections self-consistently using the linearized Boltzmann equation in contrast to the commonly used democratic Grad ansatz which ignores differences in particle scattering rates. The effect on harmonic flow coefficients should be detectable in identified particle $v_2$ and $v_4$ in heavy-ion experiments. While this study suggests that the method used to calculate the corrections will impact particle observables, it is by no means the final word on the effects. Improvements are being made to use a realistic viscous hydrodynamics code along with a full list of energy-dependent cross sections of hadrons for future work.

\ack
This work was supported by the U.S. Department of Energy, Office of Science, under grants DE-AC02-98CH10886 [RIKEN BNL] and DE-PS02-09ER41665.

\section*{References}

\end{document}